\documentclass{ws-p9-75x6-50}

\def\lsim{\lower.5ex\hbox{$\; \buildrel < \over \sim \;$}}
\def\gsim{\lower.5ex\hbox{$\; \buildrel > \over \sim \;$}}

\begin{document}

\title{Fluid flow and inertial forces in black hole space times}

\author{Banibrata Mukhopadhyay \& A. R. Prasanna} 
\address{Theoretical Physics Group, Physical Research Laboratory,
Navrangpura, Ahmedabad-380009, India
}

\maketitle

\abstracts{
We discuss the nature of the radial and azimuthal components of centrifugal
force associated with fluid flows in the equatorial plane of black hole space times.
The equations of motion are solved for the radial and azimuthal components
of the 3-velocity $V^i$ which are then used in evaluating the nature of
the various components of inertial accelerations. It is shown that the
reversal of centrifugal force is governed mainly by the dominance
of the azimuthal velocity and the reversal occurs for $r$, mostly
at $2m\lsim r \lsim 3m$, depending upon the boundary condition.
}

\centerline{\large International Journal of Modern Physics A }

\section{Introduction}

Most of the discussion on fluid flows in astrophysical context are Newtonian \cite{bon,c64,l70}
in nature, with a few considering post
Newtonian features \cite{cn,c72a,c72b,c73}. However, if
the flows are to be considered in the context of black holes, it is only
natural to expect the effects of space time curvature to affect the flows
considered in a general relativistic framework. Such considerations may
indeed show new features, quite unfamiliar in pure Newtonian descriptions.
One of the ways to study such features is to develop a full scale general
relativistic formalism of flows around black holes. However, there is one
aspect of general relativity which yields the possibility of looking
for Newtonian features, through a partitioning of
space time in a 3+1 language. It has been shown by Abramowicz et al. \cite{a93}
that one can indeed introduce the concept of `inertial forces' in
general relativity through a covariant formalism, as applied to test particle
motion in curved space time.
Recently Prasanna \cite{p01} (hereafter refers to as Paper I), using a matching
of the 3+1 conformal splitting of space-time (optical reference geometry
\cite{a88}), with the ADM splitting \cite{york} in terms of the lapse function
$\alpha$ and shift vector $\beta^i$ has expressed the components of the
covariantly defined centrifugal and Coriolis forces acting on a
fluid element in the given curved space time to be
\begin{eqnarray}
\label{cfi}
\left(Cf\right)_i=\gamma^2\left[VV^j\partial_j\left(\frac{V_i}{V}\right)
+(V_iV^j\partial_j-V^2\partial_i)\Phi-\frac{1}{2}V^jV^k\partial_i\gamma_{jk}
\right]
\end{eqnarray}
 
\noindent and
 
\begin{eqnarray}
\label{coi}
\left(Co\right)_i=-\frac{\gamma^2}{\alpha}\left[V\beta^j\partial_j\left(
\frac{V_i}{V}\right)+V^j\partial_ig_{0j}-\beta^kV^j\partial_i\gamma_{kj}\right]
\end{eqnarray}
 
\noindent where,
 
\begin{eqnarray}
\label{v}
V^2=\gamma_{ij}V^iV^j \,\,\,\,\, {\rm and}\,\,\,\,\, \gamma=\frac{1}{\sqrt{1-V^2}}
\end{eqnarray}
 
\begin{eqnarray}
\label{vi}
V^i=\frac{\left(\frac{u^i}{u^t}+\beta^i\right)}{\alpha}.
\end{eqnarray}
 
\noindent Here, $u^\mu=(u^t,u^i)$ is the four velocity of the fluid element and $\Phi$
is the gravitational potential of the back ground space time. Further it was
shown in Paper I \cite{p01} that in static space time for the case of pure dust fluid
(pressure zero) the centrifugal force would show reversal only when $V^\phi$
is non zero whereas when both $V^r$ and $V^\phi$ are non zero there is
no reversal. It is known that generally in most of the
stationary flows the radial component of the velocity vector is non zero and
one should see what happens to the force components if the pressure is
non zero and consequently the gradient  of pressure also comes into action.
It is possible that the gradient of pressure may balance the force due to
radial velocity and thereby allow the azimuthal component of the velocity
to dominate and consequently affect the centrifugal force.
 
With this in mind, we now consider the study of inertial forces for general
flows with pressure non zero and with different equations of state. We shall
in this consider the flows essentially confined to the equatorial plane $\theta=\pi/2$
and evaluate the inertial force components, after solving for the velocity
fields from the standard equation of motion on a given background geometry.

\section{Formalism}

The general equations of motion for a perfect fluid on curved space time
are given by: \\
\begin{equation}
\left( \rho + P \right) u^\mu_{\; \; \; ; \nu} u^\nu + h^{\mu\nu} P_\nu = 0
\end{equation}
Using the definition of 3-velocity $V^i$, one can write explicitly the equation of
continuity
\begin{eqnarray}
&(\rho +P) \left( u^t\right)^2 \left[ \frac{\partial}{\partial x^i} \left( \alpha V^i -
\beta^i \right) + \Gamma^\nu_{o\nu} + \Gamma^\nu_{j\nu} \left( \alpha V^j -
\beta^i \right) - \Gamma^o_{oo} \right.\\[16pt] \nonumber &
\left. - 2 \Gamma^o_{oj} \left( \alpha V^j - \beta^j \right) - \Gamma^o_{jk} \left( \alpha
V^j - \beta^j \right) \left( \alpha V^k - \beta^k \right) \right]\\[16pt] \nonumber &+ \left( u^t
\right)^2 \left[ \frac{\partial}{\partial t} \left( \rho - P \right) + \left( \alpha V^j - \beta^j
\right) \frac{\partial}{\partial x^j} \left( \rho - P \right) \right] = g^{o\nu}
\frac{\partial P}{\partial x^\nu}
\end{eqnarray}
 and the momentum equations
\begin{eqnarray}
&(\rho +P) \left( u^t\right)^2 \left[ \frac{\partial}{\partial t} \left( \alpha V^i -
\beta^i \right)+ \left( \alpha V^j - \beta^j \right) \frac{\partial}{\partial x^j} \left( \alpha V^i
- \beta^i \right)\right. \\[16pt] \nonumber &+ \Gamma^i_{oo} - \Gamma^o_{oo} \left(
\alpha V^i - \beta^i \right) + 2 \left( \alpha V^j - \beta^j \right) \left(
\Gamma^i_{oj} - \Gamma^o_{oj} \left( \alpha V^i - \beta^i \right) \right)\\[16pt] \nonumber
&\left. \left( \alpha V^j - \beta^j \right) \left( \alpha V^k - \beta^k \right) \left(
\Gamma^i_{jk} - \Gamma^o_{jk} \left( \alpha V^i - \beta^i \right) \right) \right]
\\[16pt] &= \left( \left( \alpha V^i - \beta^i \right) g^{o\nu} - g^{i\nu} \right)
\frac{\partial P}{\partial x^\nu}.
\end{eqnarray}
 
We shall presently consider
these on a stationary back ground as specified by the Kerr geometry
in Boyer-Lindquist frame:
 
\begin{eqnarray}
\label{metric}
ds^2=-\frac{(\Delta-a^2sin^2\theta)}{\rho^2}dt^2-
4a\frac{m r sin^2\theta}{\rho^2} dtd\phi+\frac{\rho^2}{\Delta}dr^2+
\rho^2d\theta^2+\frac{\Sigma}{\rho^2}sin^2\theta d\phi^2
\end{eqnarray}

\noindent with $\Delta=r^2-2mr+a^2$, $\rho^2=r^2+a^2cos^2\theta$, $\Sigma=(r^2+a^2)^2-
a^2\Delta sin^2\theta$.
 
The various components of the shift vectors $\beta_i$ and the 3-metric
$\gamma_{ij}$, alongwith the lapse function $\alpha$ are given by
(on the plane $\theta=\pi/2$),
 
\begin{eqnarray}
\label{alp}
\alpha^2=\frac{r^2+a^2-2mr}{r^2+a^2+2ma^2/r},
\end{eqnarray}
\begin{eqnarray}
\label{betai}
\beta_i=\left(0,0,0,-2am/r\right),
\end{eqnarray}
 
\begin{eqnarray}
\label{gama}
\gamma_{rr}=\left(1-\frac{2m}{r}+\frac{a^2}{r^2}\right)^{-1}\,\,\,\,\,\,
{\rm and} \,\,\,\,\,\, \gamma_{\phi\phi}=r^2+a^2+\frac{2ma^2}{r}
\end{eqnarray}
 
\noindent while the gravitational potential $\Phi$ is given as
 
\begin{eqnarray}
\label{phi}
\Phi=-\frac{1}{2}ln\left[\frac{r^3+a^2r-2mr^2}{r^3+a^2r+2ma^2}\right].
\end{eqnarray}

Using these in the general equations referred to, alongwith the components
of connection coefficients for the metric (\ref{metric}) one gets,\\
the momentum equation:
 
\begin{eqnarray}
\label{eqn1}
(P+\rho)[{\bar u^r}\frac{d{\bar u^r}}{dr}+\left\{\frac{a^2-mr}{\Delta r}-
\frac{2m(a^2+r^2)}{\Delta r^2}\right\}({\bar u^r})^2+\frac{2am}{\Delta r^2}
(a^2+3r^2)({\bar u^r})^2{\bar u^\phi} \\ \nonumber
+\frac{\Delta}{r^4}(ma^2-r^3)({\bar u^\phi})^2
-\frac{2ma\Delta}{r^4}{\bar u^\phi} +\frac{\Delta m}{r^4}] \\ \nonumber
=\frac{\Delta}{r^2}\left[-\frac{\Delta-a^2}{r^2}+\frac{r^2}{\Delta}
({\bar u^r})^2
-\frac{4am}{r}{\bar u^\phi}+\frac{(r^2+a^2)^2-a^2\Delta}{r^2}
({\bar u^\phi})^2\right]\frac{dP}{dr},
\end{eqnarray}
 
\begin{eqnarray}
\label{eqn2}
\frac{d{\bar u^\phi}}{dr}+\frac{2am}{\Delta r^2}(a^2+3 r^2)({\bar u^\phi})^2
-\frac{2}{\Delta r^2}(2ma^2+r^2(3m-r)){\bar u^\phi}+\frac{2ma}{\Delta r^2}=0
\end{eqnarray}
 
\noindent and the equation of continuity:
 
\begin{eqnarray}
\label{eqn3}
\frac{1}{{\bar u^r}}\frac{d{\bar u^r}}{dr}+
\frac{2am}{\Delta r^2}(a^2+3r^2){\bar u^\phi}-\frac{m}{\Delta r^2}(a^2+r^2)
+\frac{1}{r\Delta}(a^2-mr)\\ \nonumber
-\frac{1}{\Delta r^2}(ma^2+r^2(2m-r))
=\frac{1}{(P+\rho)}\frac{d}{dr}(P-\rho),
\end{eqnarray}
 
\noindent where,
 
\begin{eqnarray}
\label{ubar}
{\bar u^i}=\frac{u^i}{u^t}=\alpha V^i-\beta^i.
\end{eqnarray}

We shall consider the flows with different equation of state and look
at the behaviour of the force components. For the present we are restricting
ourselves to the case of flows on the equatorial plane ($\theta=\pi/2$)
of the central gravitating source, a condition which has been incorporated in the
equations above.

\section{ Solutions }
 
\subsection*{Case (1), P=0; Dusty Fluid}
 
As this refers to the case when streamlines are geodesics, the results are same as
obtained in the case of test particles, considered in `Paper I'.

\subsection*{Case (2), $\rho$=constant}
 
\noindent {\large\bf (a) Schwarzschild space time}\\ We shall first consider the
static case by putting $a=0$ and consequently have the equations:
 
\begin{eqnarray}
\label{srad}
\alpha V^r\frac{d}{dr}(\alpha V^r)+\frac{m}{r^2}\left(1-\frac{2m}{r}\right)
-\frac{3m}{r^2}\left(1-\frac{2m}{r}\right)^{-1}(\alpha V^r)^2-(r-2m)
(\alpha V^\phi)^2\\ \nonumber
=-\frac{1}{(u^t)^2}\left(1-\frac{2m}{r}\right)\frac{1}{(P+
\rho)}\frac{dP}{dr},
\end{eqnarray}
 
\begin{eqnarray}
\label{saz}
\frac{d}{dr}(\alpha V^\phi)+\frac{2}{r}\left(1-\frac{2m}{r}\right)^{-1}
\left(1-\frac{3m}{r}\right)\alpha V^\phi=0
\end{eqnarray}
 
\noindent and the equation of continuity:
 
\begin{eqnarray}
\label{scon}
\frac{d}{dr}(\alpha V^r)+\frac{2}{r}\left(1-\frac{2m}{r}\right)^{-1}
\left(1-\frac{3m}{r}\right)\alpha V^r=\frac{\alpha V^r}{(P+\rho)}\frac{dP}{dr}.
\end{eqnarray}
 
\noindent Thus one finds $\alpha V^\phi=\frac{l}{r^2}\left(1-\frac{2m}{r}\right)$ and
$\alpha V^r$ satisfies the equation
 
\begin{eqnarray}
\label{eee}
-\left(1-\frac{2m}{r}\right)^2\left[1-\frac{l^2}{r^2}\left(1-
\frac{2m}{r}\right)\right]\frac{1}{\alpha V^r}\frac{d}{dr}(\alpha V^r)
+\frac{2}{r}\left(1-\frac{3m}{2r}\right)\left(1-\frac{2m}{r}\right)^{-1}
(\alpha V^r)^2 \\ \nonumber
+\left(1-\frac{2m}{r}\right)\left[-\frac{2}{r}+\frac{5m}{r^2}+
\frac{l^2}{r^3}\left(1-\frac{2m}{r}\right)\left(3-\frac{8m}{r}\right)\right]=0.
\end{eqnarray}
 
\noindent Solving which one can get
 
\begin{eqnarray}
\label{}
(\alpha V^r)^2=\frac{\left(1-\frac{2m}{r}\right)^2\left[1-\frac{l^2}{r^2}
\left(1-\frac{2m}{r}\right)\right]}{\left[1+c_1 r^4 \left(1-\frac{2m}{r}\right)
\right]},
\end{eqnarray}
 
\noindent where $c_1$ is the constant of integration to be chosen appropriately.
The components of centrifugal force are then given by
 
\begin{eqnarray}
\label{scenr}
(Cf)_r=\frac{l^2\left[(2m-r)(c_1 r^2 (6m^2 -5rm+r^2)-1)l^2+(3m-2r)r^2\right]}
{r^3(r^3+l^2(2m-r))(c_1 l^2 (r-2m)^2+1)}
\end{eqnarray}
 
\begin{eqnarray}
\label{scenp}
(Cf)_\phi=-\frac{l\left[(2m-r)(c_1 r^2 (6m^2-5rm+r^2)-1)l^2+(3m-2r)r^2\right]}
{r^4\sqrt{\frac{r^3+l^2(2m-r)}{r^3}}\sqrt{c_1(r-2m)r^3+1}(c_1 l^2 (r-2m)^2+1)}.
\end{eqnarray}
 
\noindent It should be noted at the outset that for $l=3\sqrt{3}m$, at $r=3m$, the
total velocity $V$ tends to $1$, the velocity of light and thus makes the
Lorentz factor infinite. This feature is independent of the constant
$c_1$ as may be seen from the expression for $V$ ($=[1+c_1 l^2 (r-2m)^2]/[1
+c_1 r^3 (r-2m)]$). Thus, restricting the value of $l<3\sqrt{3}m$, it may be
seen that the centrifugal reversal occurs for two values of $r$ one close
to $r=3m$ and other close to the horizon $r=2m$, depending upon the choice
of $l$ and $c_1$. For very low values of $l$ and $c_1$ the real roots may
all lie within the horizon (e.g. for $l=1$ and $c_1=6$, $r_1=1.73$,
$r_2=0.175$, $r_3=-0.153$). In principle there are five zeros for the quintic
equation obtainable by setting $Cf=0$, but only two of them lie outside
the horizon. For a fixed $l$ as $c_1$ is increased, the outer most root stays
fixed at $r=3m$, while the other one moves closer and closer to the horizon.
Figures 1 gives the plots for the components of the centrifugal force for
different values of $l$ and $c_1$.
 
\begin{figure}
\vbox{
\vskip -0.5cm
\hskip 0.0cm
\centerline{
\psfig{figure=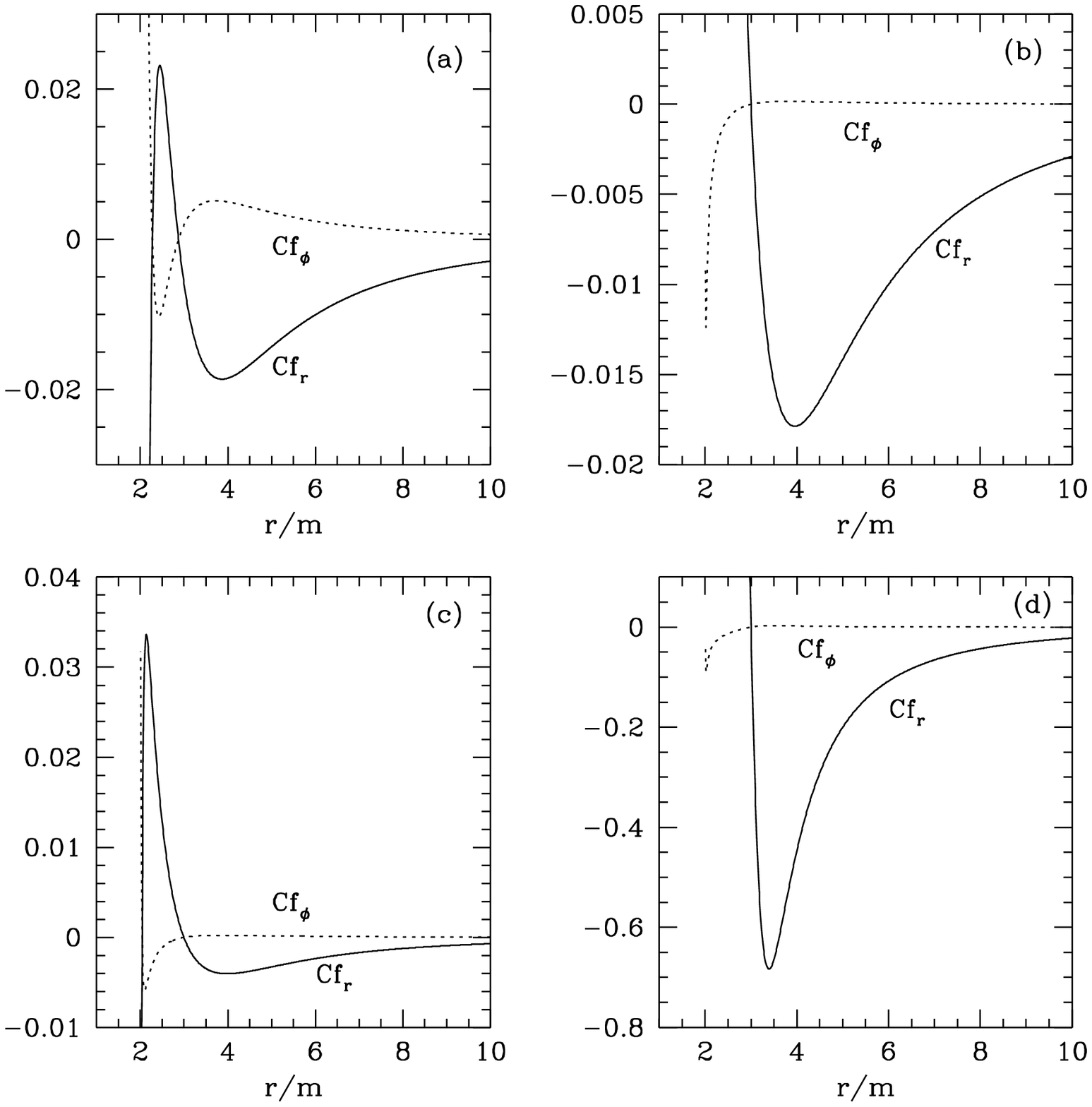,height=10truecm,width=12truecm,angle=0}}}
\vspace{-0.0cm}
\noindent {\small {\bf Fig. 1} : $\rho$=constant, Schwarzschild case.
It shows the variation of $Cf_r$ and $Cf_\phi$
as functions of $r/m$ for (a) $l=2$ and $c_1=6$, (b) $l=2$ and $c_1=6000$,
(c) $l=1$ and $c_1=600$, (d) $l=5$ and $c_1=600$.
}
\end{figure}

\noindent { \large\bf (b) Kerr space time}\\ The equations of motion for this
background with $\rho={\rm constant}$ may be obtained from the general
equations (\ref{eqn1}), (\ref{eqn2}) and (\ref{eqn3}). Though the equation
for the azimuthal velocity is decoupled from the other two, the equations
are not tractable to solve analytically. Thus integrating them numerically
with the appropriate boundary conditions one can obtain the profiles for
the velocity, using which the profiles for the force components may be
obtained. Figure 2 gives the plots for the various profiles for different
values of the Kerr parameter $a$. As may be seen, there is a definite
reversal of both components of the centrifugal force in each case at
locations close to $r\le 3m$ (see Table 1). The pressure profile shows a monotonic
increase close to the central compact object.
 
\vskip0.5cm
\begin{figure}
\vbox{
\vskip -0.5cm
\hskip 0.0cm
\centerline{
\psfig{figure=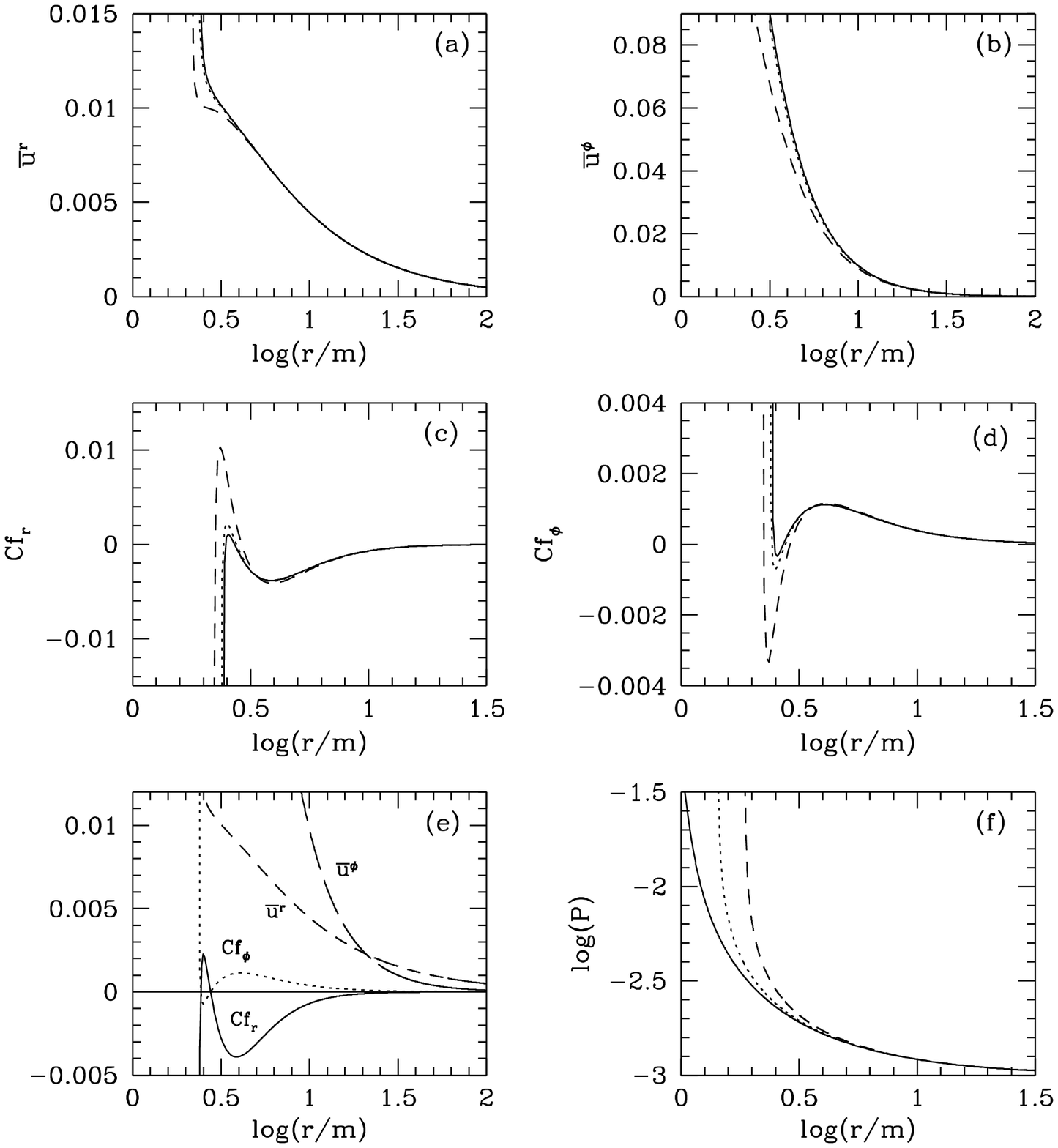,height=12truecm,width=14truecm,angle=0}}}
\vspace{-0.0cm}
\noindent {\small {\bf Fig. 2} : $\rho$=constant, Kerr case.
It shows the variation of (a) ${\bar u^r}$, (b) ${\bar u^\phi}$, 
(c) $Cf_r$, (d) $Cf_\phi$ as functions of radial
distance in unit of mass of the black hole ($r/m$). Solid, dotted and
dashed curves indicate the results for Kerr parameter $a=1$, $0.9$ and
$0.5$ respectively. (e) Variation of ${\bar u^r}$, ${\bar u^\phi}$,
$Cf_r$ and $Cf_\phi$ as a function of $r/m$ for $a=0.9$ which shows
the relation of the centrifugal reversal location with the location
where ${\bar u}^\phi$ crosses over ${\bar u}^r$.
(f) Variation of pressure $P$ as a function of radial distance $r/m$.
}
\end{figure}
 
\subsection*{ Case (3), $\rho=P$; Isothermal Fluid}
 
\noindent {\large\bf (a) Schwarzschild space time}\\ In the extreme case of isothermal
equation of state the equations of motion on the Schwarzschild background
reduce to
\begin{eqnarray}
\label{srad}
2P(u^t)^2[\alpha V^r\frac{d}{dr}(\alpha V^r)+\frac{m}{r^2}\left(1-
\frac{2m}{r}\right)-\frac{3m}{r^2}\left(1-\frac{2m}{r}\right)^{-1}
(\alpha V^r)^2 \\ \nonumber
-(r-2m)(\alpha V^\phi)^2]=-\left(1-\frac{2m}{r}\right)\frac{dP}{dr},
\end{eqnarray}
 
\begin{eqnarray}
\label{saz}
\frac{d}{dr}(\alpha V^\phi)+\frac{2}{r}\left(1-\frac{2m}{r}\right)^{-1}
\left(1-\frac{3m}{r}\right)\alpha V^\phi=0,
\end{eqnarray}
 
\begin{eqnarray}
\label{sacon}
\frac{d}{dr}(\alpha V^r)+\frac{2}{r}\left(1-\frac{2m}{r}\right)^{-1}\left(1-
\frac{3m}{r}\right)\alpha V^r=0
\end{eqnarray}
 
\noindent and
 
\begin{eqnarray}
\label{sacon}
(u^t)^{-2}=\left[\left(1-\frac{2m}{r}\right)-\left(1-\frac{2m}{r}\right)^{-1}
(\alpha V^r)^2-r^2 (\alpha V^\phi)^2\right].
\end{eqnarray}
 
\noindent It is seen from (\ref{saz}) that the angular velocity shows the same behaviour
as in Case 2(a) with $\alpha V^\phi=\frac{l}{r^2}\left(1-\frac{2m}{r}\right)$.
The radial velocity comes out constrained from the continuity equation equation
to be $\alpha V^r=\frac{v_1}{r^2}\left(1-\frac{2m}{r}\right)$, $v_1$ being constant
of integration, which by dimensional considerations and also due to the consistency
with the equation of continuity can be identified with the ratio ${\dot M}/\rho$,
$\dot M$ being the accretion rate and $\rho$ the density. With these, the components
of centrifugal force turn out to be
 
\begin{eqnarray}
\label{sperhcr}
(Cf)_r=-\frac{\gamma^2 l^2}{V^2 r^5}\left(1-\frac{2m}{r}\right)\left[\frac{2v_1^2}{r^2}
+l^2\left(1-\frac{3m}{r}\right)\right],
\end{eqnarray}
 
\begin{eqnarray}
\label{sperhcr}
(Cf)_\phi=\frac{\gamma^2 l v_1}{V^2 r^3}\left(1-\frac{2m}{r}\right)\left[\frac{2v_1^2}{r^2}
+l^2\left(1-\frac{3m}{r}\right)\right]
\end{eqnarray}
 
\noindent with $\gamma^2=(1-V^2)^{-1}$ and $V^2=V^rV_r+V^\phi V_\phi$.
From these expressions it is again clear that these do not change sign for any $r>3m$.
However, for $r<3m$, both components of the centrifugal force change sign for
$r=\frac{3m}{2}\left(1\pm\sqrt{1-\frac{8v_1^2}{9l^2m^2}}\right)$. Obviously the reversal
depends upon the reality of $r$, which demands $\frac{{\dot M}}{\rho}\le\frac{3lm}{2\sqrt{2}}$.
Figure 3 shows the profiles of velocity and centrifugal forces for four different values of
the ratio $\frac{lm}{v_1}$. For $\frac{lm}{v_1}\le 1$ the root lies inside the horizon
$r=2m$ and then the reversal can not be seen. On the other hand for $\frac{lm}{v_1}>1$
the root lies outside the horizon $r=2m$ but at the location $<3m$.
 
\vskip0.5cm
\begin{figure}
\vbox{
\vskip -0.5cm
\hskip 0.0cm
\centerline{
\psfig{figure=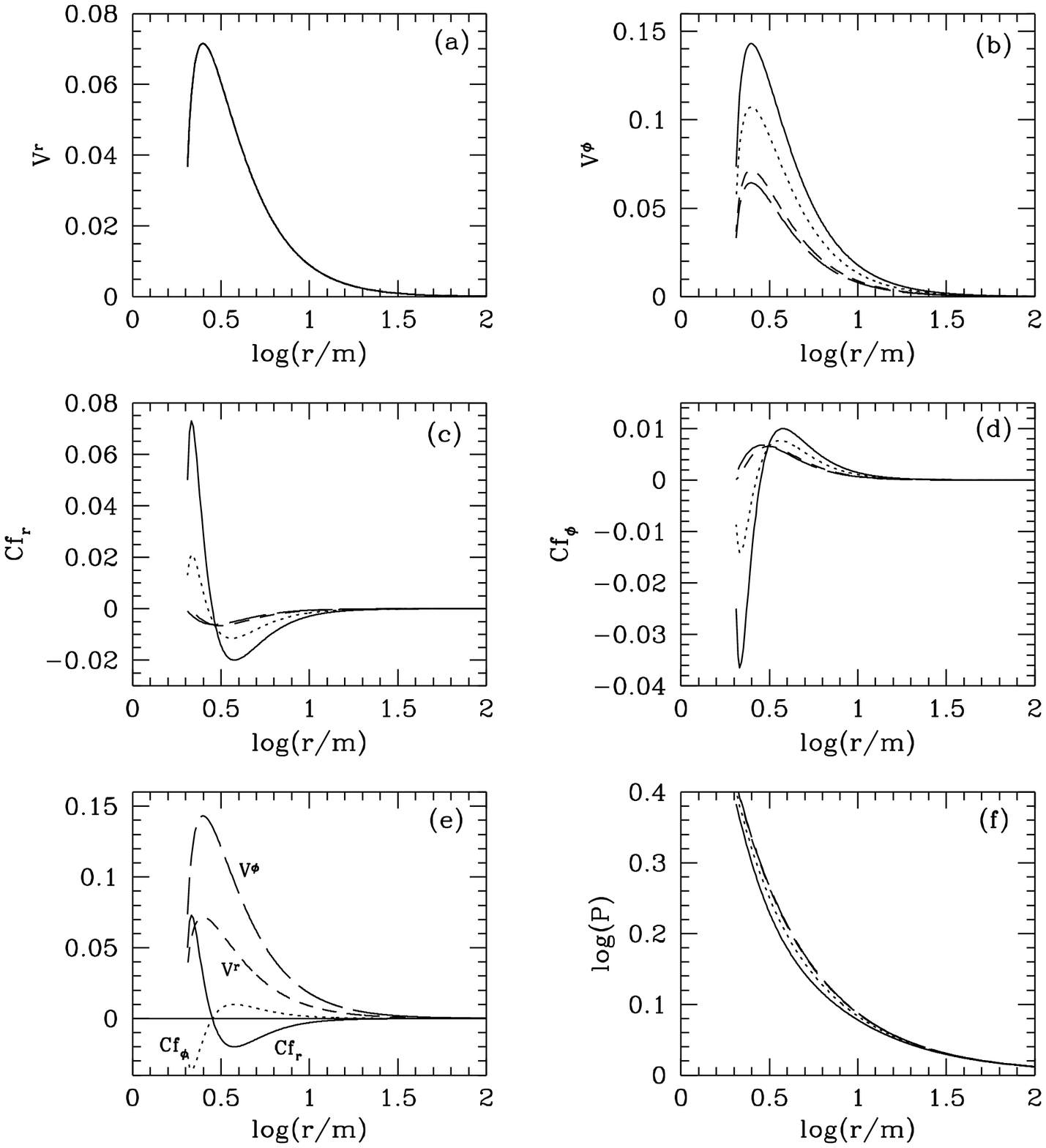,height=10truecm,width=12truecm,angle=0}}}
\vspace{-0.0cm}
\noindent {\small {\bf Fig. 3} : $P=\rho$, Schwarzschild case.
It shows the variation of (a) $V^r$,
(b) $V^\phi$, (c) $Cf_r$, (d) $Cf_\phi$ as functions of radial
distance in unit of mass of the black hole ($r/m$). Solid, dotted,
dashed and long-dashed curves indicate the results for $l/v_1=2$, $1.5$, $1$ and
$0.9$ respectively. (e) Variation of $V^r$, $V^\phi$,
$Cf_r$ and $Cf_\phi$ as functions of $r/m$ for $l/v_1=2$, which shows
the dominance of azimuthal velocity over radial velocity and
the reversal of centrifugal forces. (f) Variation of pressure $P$ as a function of $r/m$.
}
\end{figure}
 
\noindent {\large\bf (b) Kerr space time}\\ In the case of Kerr background, again the equations
are integrated numerically and the profiles presented in Fig. 4. The plots give again the
velocity and force components alongwith the pressure profile for different values of
$a$. Here again as in the case of $\rho$ constant, the centrifugal force changes sign at a radius
$r$ between $2m$ and $3m$, the exact location of the zero depending upon the
boundary condition (see Table 2).
 
\vskip0.5cm
\begin{figure}
\vbox{
\vskip -0.5cm
\hskip 0.0cm
\centerline{
\psfig{figure=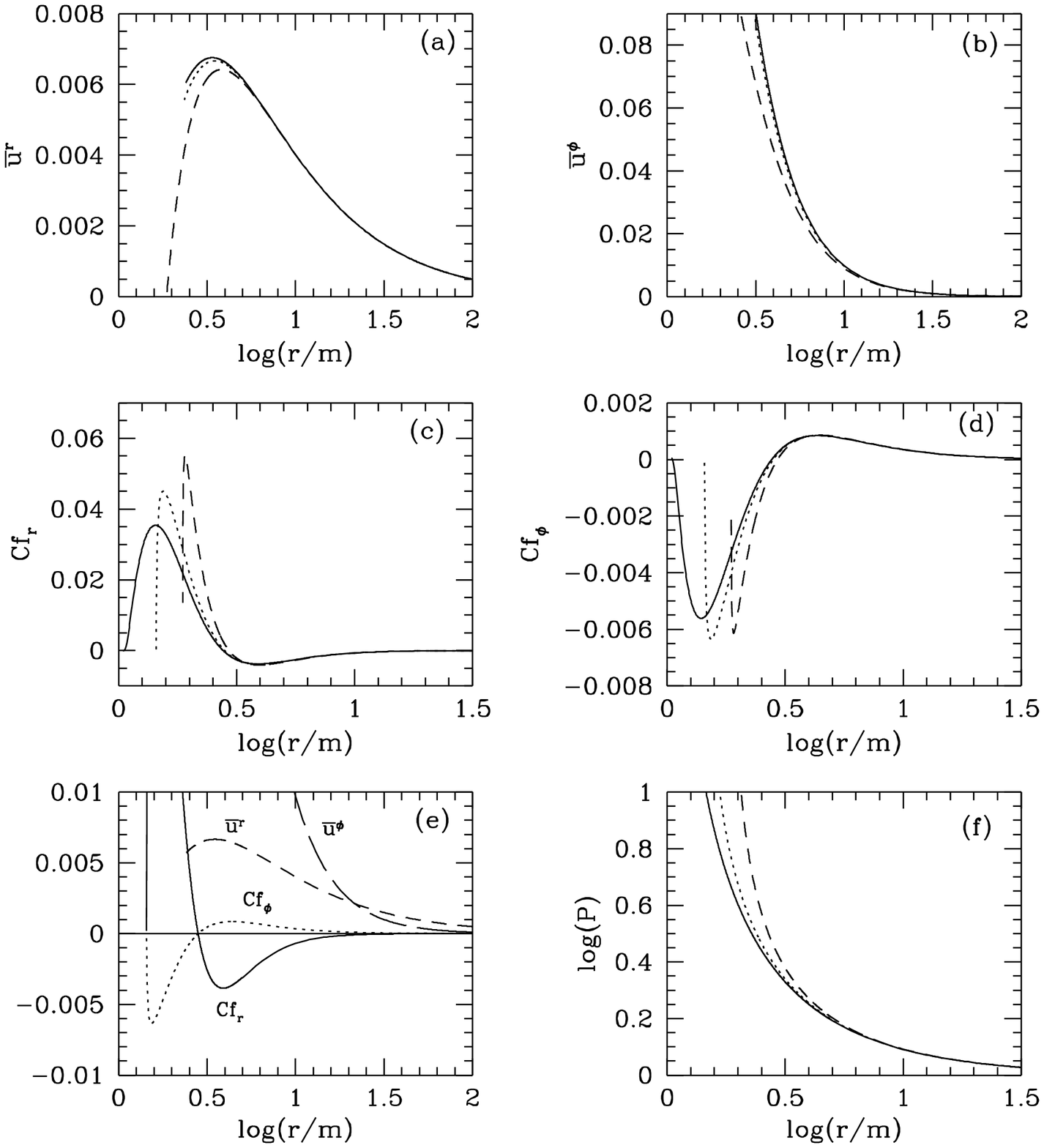,height=10truecm,width=12truecm,angle=0}}}
\vspace{-0.0cm}
\noindent {\small {\bf Fig. 4} : $P=\rho$, Kerr case. Plots are
similar to the ones as Fig. 3.
}
\end{figure}
 
\subsection*{Case (4), $P=\rho^\Gamma$; $\Gamma$ being the gas constant}
 
\noindent The adiabatic equation of state yields for the static space time, the fluid
equations which are coupled for $V^r$ and $P$, while the azimuthal component of the velocity
again has the same profile $V^\phi=\frac{l}{r^2}\left(1-\frac{2m}{r}\right)^{1/2}$.
The equations for $V^r$ and $P$ are solved numerically for different values of $\Gamma$ and
the results are presented in Fig. 5 for three different values of $\Gamma$; $5/3$, $4/3$ and
$1.0001$. The radial velocity profile shows different slopes for different values
of the gas constant
 
\vskip0.5cm
\begin{figure}
\vbox{
\vskip -0.5cm
\hskip 0.0cm
\centerline{
\psfig{figure=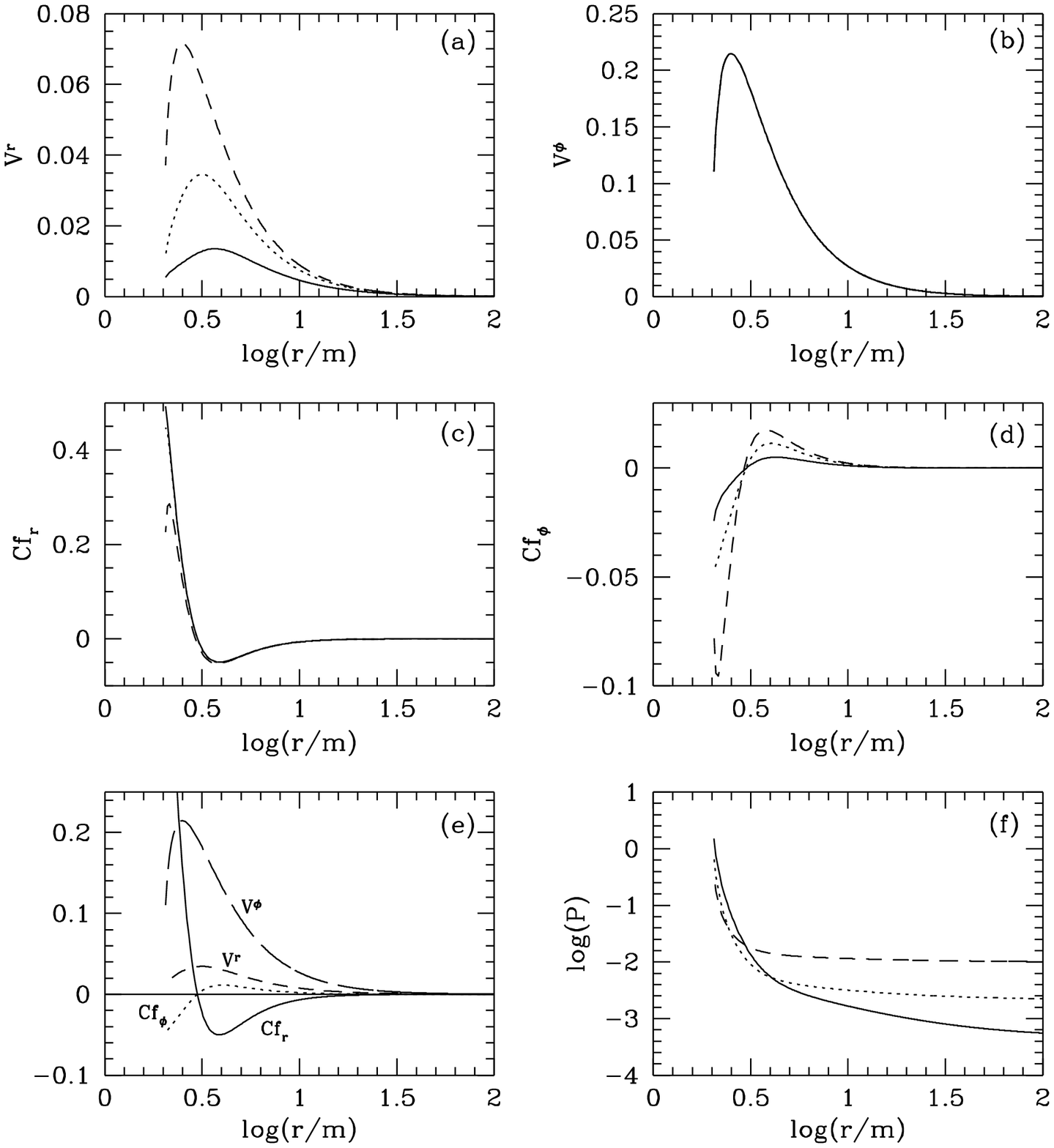,height=10truecm,width=12truecm,angle=0}}}
\vspace{-0.0cm}
\noindent {\small {\bf Fig. 5} : Schwarzschild case for adiabatic equation of state.
It shows the variation of (a) $V^r$,
(b) $V^\phi$, (c) $Cf_r$, (d) $Cf_\phi$, (f) $P$ as functions of radial
distance in unit of mass of the black hole ($r/m$). Solid, dotted and
dashed curves indicate the results for gas constant $\Gamma=5/3$, $4/3$, $1.0001$
respectively. (e) Variation of $V^r$, $V^\phi$,
$Cf_r$ and $Cf_\phi$ as functions of $r/m$ for $\Gamma=4/3$ which indicates
the dominance of azimuthal velocity over radial velocity  and
the reversal of centrifugal forces.
}
\end{figure}
 
\noindent which behaviour is reflected in the profiles for pressure and density too. The azimuthal
velocity profile is independent of $\Gamma$. Again
the centrifugal force shows reversal in the region $2m<r<3m$, in accordance with
other cases. Though both components of the velocity show a decreasing trend, calculations
show that the total velocity $V$ tends toward $c (=1)$ as $r\rightarrow 2m$, which is in
accordance with expectations.

Figures 6 and 7 show the profiles for the same physical quantities in the case
of Kerr background for the two distinct cases, $\Gamma=4/3$ and $\Gamma=5/3$ and for different values
of the Kerr parameters `$a$'. The behaviour of various quantities are just similar, again
showing the feature of centrifugal reversal in the region $2m<r\le3m$ which is outside the
ergo region.
 
\vskip0.5cm
\begin{figure}
\vbox{
\vskip -0.5cm
\hskip 0.0cm
\centerline{
\psfig{figure=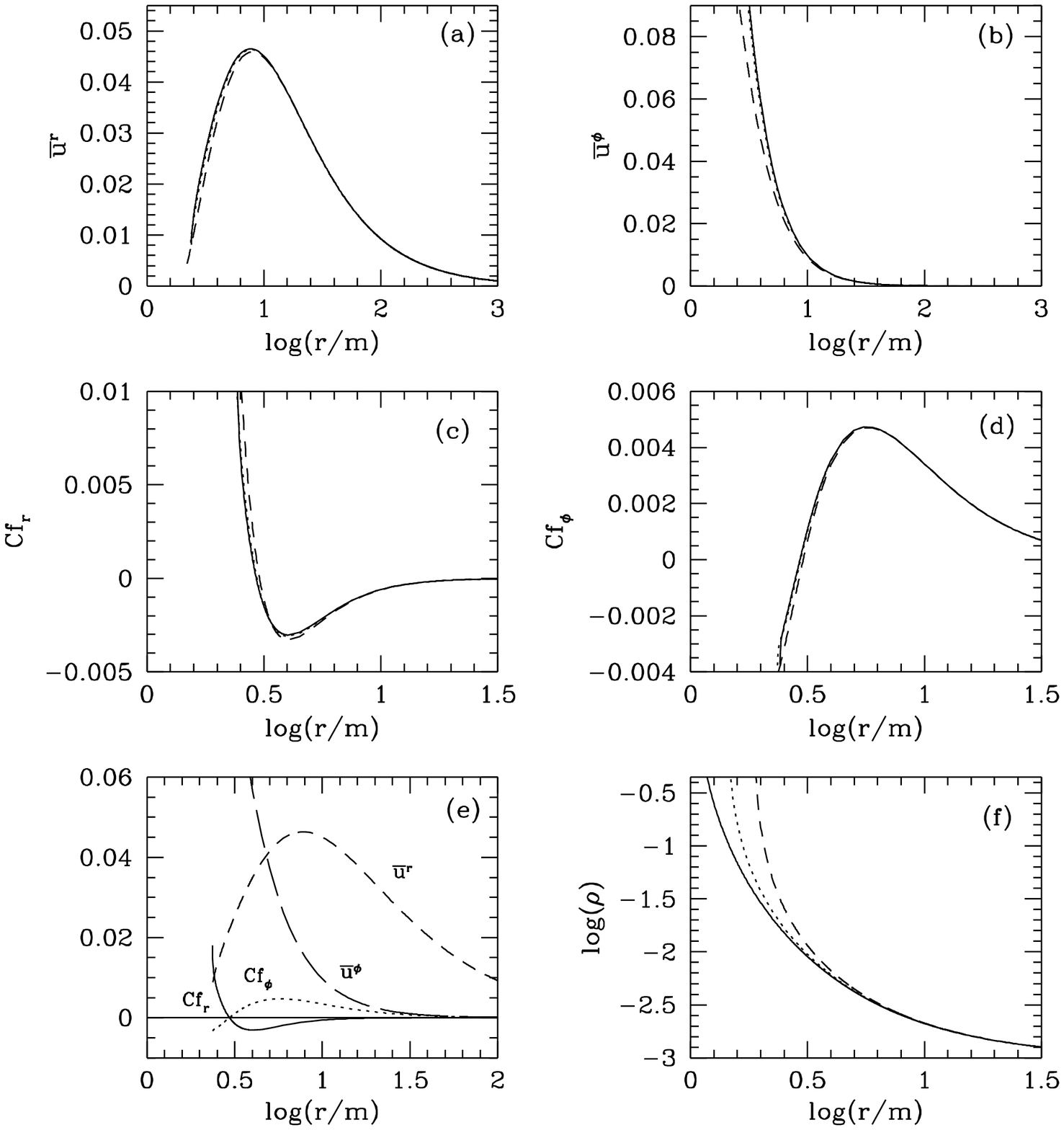,height=11truecm,width=13truecm,angle=0}}}
\vspace{-0.0cm}
\noindent {\small {\bf Fig. 6} : Kerr case for adiabatic equation of state with
$\Gamma=4/3$. Plots are similar to the ones as Fig. 3.
}
\end{figure}
 
\begin{figure}
\vbox{
\vskip -0.5cm
\hskip 0.0cm
\centerline{
\psfig{figure=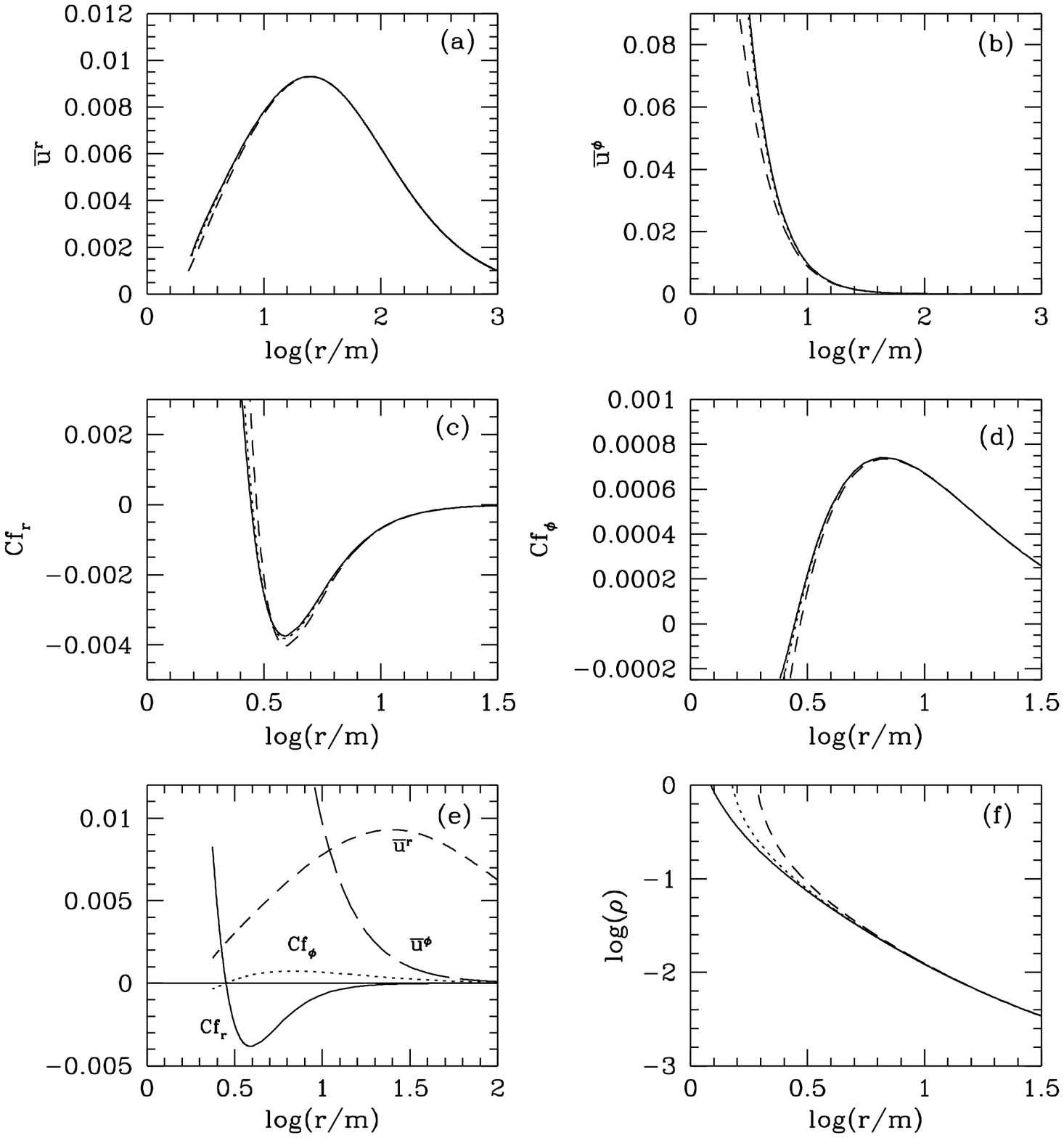,height=12truecm,width=14truecm,angle=0}}}
\vspace{-0.0cm}
\noindent {\small {\bf Fig. 7} : Kerr case for adiabatic equation of state with
$\Gamma=5/3$. Plots are similar to the ones as Fig. 3.
}
\end{figure}
 
\section{ Discussion}
 
\noindent We shall first consider the azimuthal component of the centrifugal force in the Schwarzschild
geometry for the case of dusty fluid, which turns out to be
 
\begin{eqnarray}
\label{scfphp0}
(Cf)_\phi=\frac{l}{2r}\left[\frac{2mc^2}{r}-\frac{l^2}{r^2}\left(1-
\frac{2m}{r}\right)\right]^{1/2}
\end{eqnarray}
 
\noindent and shows no reversal just as the radial component as considered earlier in 
Paper I \cite{p01}.
On the other hand when we consider the Kerr background the force components do show a reversal
even when $P=0$. Obviously the extra contribution towards angular velocity
comes from the dragging of inertial frames, an effect that dominates as
$r\rightarrow 2m$ the ergo surface. Thus one finds the reversal of the
centrifugal force to occur close to the location where ${\bar u}^\phi$
crosses ${\bar u}^r$.
 
When the density is constant with pressure non zero, the analytical expressions are obtained for
the components of the centrifugal force in the static case, which show that though in principle
there can be five roots for the equation $(Cf)=0$, only two of them are outside the event
horizon $r=2m$. However the realisation of these two roots also depend upon the condition that
$l$ be $\le 3\sqrt{3}m$. For $l=3\sqrt{3}m$, $V\rightarrow c$, as $r\rightarrow 3m$, which indeed is interesting, for
the minimum impact parameter for a zero rest mass particle in the Schwarzschild space time is $3\sqrt{3} m$.For this value of $l$ the centrifugal force is infinite at $r=3m$
since the Lorentz factor goes to infinity. However, for other values of $l$ the components
have two zeros, one close to $3m$ and the other close to the horizon. The occurrence of the two
zeros in the region outside the horizon does indeed depend upon the choice of the constants
$l$ and $c_1$. The reversal for a given $l$, if occur at $r=3m$ remains unchanged for all
$c_1$ whereas the one nearer the horizon gets closer to the horizon with increasing values
of $c_1$. It is also important to note that for no value of $l$ and $c_1$ the reversal occurs for
$r>3m$.
 
In the case of Kerr space time, where the equations are integrated numerically and one needs to choose the
boundary conditions appropriately, it may be seen that the location of the reversal does change
with the relative magnitudes of the velocity components at infinity. Table 1 gives the locations
of the reversal for different values of `$a$' with corresponding boundary conditions for
$\rho$=constant case. As may be seen in the first case, ${\bar u^\phi_o}>{\bar u^r_o}$ (the subscript
`$o$' refers to the values at boundary from where the integration starts), and this is maintained throughout the
flow with the azimuthal component dominating, the centrifugal force reversal occurs at the same
location as for a test particle \cite{p01,ip93}. On the other hand in the second
case though ${\bar u^r_o}>{\bar u^\phi_o}$, as the flow approaches the black hole the ${\bar u^\phi}$
takes over ${\bar u^r}$ at around $r=20m$ leading to the reversal of centrifugal force at the
location for example $r=2.783m$ for $a=0.9$ (Fig. 2e), a location slightly closer to the black
hole than the reversal radius  of $2.822m$ for the test particle case. The other root seems to
occur well inside the horizon and thus of no consequence.
\vskip0.5cm
\newpage
{\centerline{\large Table 1}}
{\centerline{\large $\rho$=constant}}
\begin{center}
{
\vbox{
\begin{tabular}{llllllllllllllllllllllllllllllll}
\hline
\hline
$a$ & & & & $0$ & & $0.5$ & & $0.9$ & & $1$ & & & & Boundary Values  \\
\hline
\hline
 & & & & & & & & & & & & & & ${\bar u^r_o}=10^{-8}$  \\
$R$  & & & & $3.0$ & & $2.944$ & & $2.822$  & & $2.783$  & & & & ${\bar u^\phi_o}=10^{-6}$  \\
& & & & & & & & & & & &  & &  $\rho_o=10^{-5}$   \\
\hline
\hline
$R_1$  & & & & $2.955$ & & $2.904$ & & $2.783$ & & $2.744$ & & & & ${\bar u^r_o}=5\times10^{-5}$   \\
$R_2$  & & & & $2.117$ & & $1.948$ & & $1.596$  & & $1.371$ & & & & ${\bar u^\phi_o}=10^{-6}$    \\
& & & & & & & & & & & & & &  $\rho_o=10^{-3}$    \\
\hline
\hline
\end{tabular}
}}
\end{center}
\vskip1.0cm
 
When we consider the isothermal equation of state ($P=\rho$) in the Schwarzschild background
one could solve the equations analytically yielding the velocity components in terms of two
constants $v_1$ and $l$. As mentioned earlier $v_1$ can be identified as the ratio ${\dot M }/\rho$
and it is seen that the expressions for the components of centrifugal force depend upon both
$v_1$ and $l$, with the reversal occurring at a value of $r=\frac{3m}{2}\left(1\pm\sqrt{1-\frac{8v_1^2}
{9l^2m^2}}\right)$. One can straight away see from this expression that the reversal would occur at
$r=3m$ for $\frac{v_1}{l}<1$, which is in accordance with earlier observation as in the case
where ${\bar u^\phi}$ would dominate over ${\bar u^r}$. On the other hand when $\frac{v_1}{l}>
\frac{3m}{2\sqrt{2}}$ or equivalently $\frac{{\dot M}}{l \rho}>\frac{3m}{2\sqrt{2}}$ then the centrifugal
will never be zero for any $r$ and thus {\it no reversal} would occur. This again confirms
the fact that when the radial velocity dominates over the angular velocity there is no reversal of
the centrifugal force. Table 2 presents the location of centrifugal reversal for the case $P=\rho$,
which distinctly shows the possibility of the reversal occurring either at the same position
as for a test particle or at locations different but inwards and closer to the black hole.
\vskip0.5cm
{\centerline{\large Table 2}}
{\centerline{\large $P=\rho$}}
\begin{center}
{
\vbox{
\begin{tabular}{llllllllllllllllllllllllllllllll}
\hline
\hline
$a$  & & & & $0$ & & $0.5$ & & $0.9$  & & $1$  & & & & Boundary Values \\
\hline
\hline
& & & & & & & & & & & & & &  ${\bar u^r_o}=10^{-8}$  \\
$R$  & & & & $3.0$ & & $2.944$ & & $2.822$ & & $2.783$  & & & & ${\bar u^\phi_o}=10^{-6}$   \\
& & & &  & &  & &  & &  & &  & & $\rho_o=10^{-5}$   \\
\hline
\hline
$R_1$ & & & & $2.992$  &  & $2.936$  & & $2.813$ & & $2.773$  & & & & ${\bar u^r_o}=5\times10^{-5}$  \\
$R_2$  & & & & No root  & & No root  & & $1.438$  & & $1.058$  & & & & ${\bar u^\phi_o}=10^{-6}$   \\
& & & &  & &  & &  & &  & & & & $\rho_o=10^{-3}$   \\
\hline
\hline
\end{tabular}
}}
\end{center}
\vskip1.0cm
In the case of Kerr black hole one set of locations are always outside the ergosurface $r=2m$,
while the other could be inside.

Coming to the case of polytropic equation of state Table 3
presents the locations of the centrifugal reversal for different values of $a$ and of gas
constant $\Gamma$ ($=1+\frac{1}{n}$, $n$ being the polytropic index). It is interesting to
see that for $\Gamma\ge 5/3$ the location of centrifugal reversal coincides with those
of a test particle, indicating the dominance of ${\bar u^\phi}$ over ${\bar u^r}$. On the
other hand for $\Gamma$, $4/3\le\Gamma<5/3$, the reversal occurs at locations slightly
higher than those corresponding to the test particle case. As may be seen from Fig. 6e and
Fig. 7e, the location of the intersection of ${\bar u^r}$ and ${\bar u^\phi}$ profiles
\vskip0.5cm
{\centerline{ \large Table 3}}
{\centerline{\large $P=\rho^\Gamma$}}
\begin{center}
{
\vbox{
\begin{tabular}{llllllllllllllllllllllllllllllllllllllllll}
\hline
\hline
$\Gamma$  & & & & $a$  & & & & $0$ & & $0.5$  & & $0.9$  & & $1$ & & &  & $n$ \\ \hline
\hline
$4/3$ & & & & $R_1$ &  & & & $3.068$ & & $3.017$ & & $2.90$ & & $2.860$ & & & &
$3$   \\
& & & & $R_2$  & & & & $2.005$  & & $1.872$ & & $1.447$ & & $1.091$ & &  & & 
   \\
\hline
\hline
$1.45$  & & & & $R_1$  & & & & $3.011$ & & $2.958$ & & $2.838$ & & $2.799$ & & & & $\sim 2.22$   \\
& &  & & $R_2$  & & & & $2.003$ & & $1.870$ & & $1.444$ & & $1.081$  & & & &
  \\
\hline
\hline
$3/2$ & & & & $R_1$ & & & & $3.005$ & & $2.950$ & & $2.830$ & & $2.790$ & & & &
$2$    \\
 & & & & $R_2$ & & & & $2.003$  & & $1.870$ & & $1.443$  & & $1.078$  & & & &
   \\
\hline
\hline
$5/3$ & & & & $R_1$ & & & & $3.0$  & & $2.945$ & & $2.823$  & & $2.783$  & & & & $1.5$    \\
 & &  & & $R_2$ & & & & $2.002$ & & $1.869$ & & $1.442$ & & $1.072$  & & & &
   \\
\hline
\hline
$2$ & & & & $R_1$ & & & & $3.0$ & & $2.945$ & & $2.823$ & & $2.783$ & & & & $1$
  \\
 & &  & & $R_2$ & & & & $2.002$ & & $1.868$ & & $1.441$ & & $1.067$  & & & &
  \\
\hline
\hline
\end{tabular}
}}
\end{center}
\vskip0.5cm
seem to have a direct bearing on the location of the centrifugal reversal. Figure 8 shows the
variation of the location of centrifugal reversal as a function of the gas parameter $\Gamma$,
for different values of the Kerr parameter $a$.
 
\begin{figure}
\vbox{
\vskip -0.5cm
\hskip 0.0cm
\centerline{
\psfig{figure=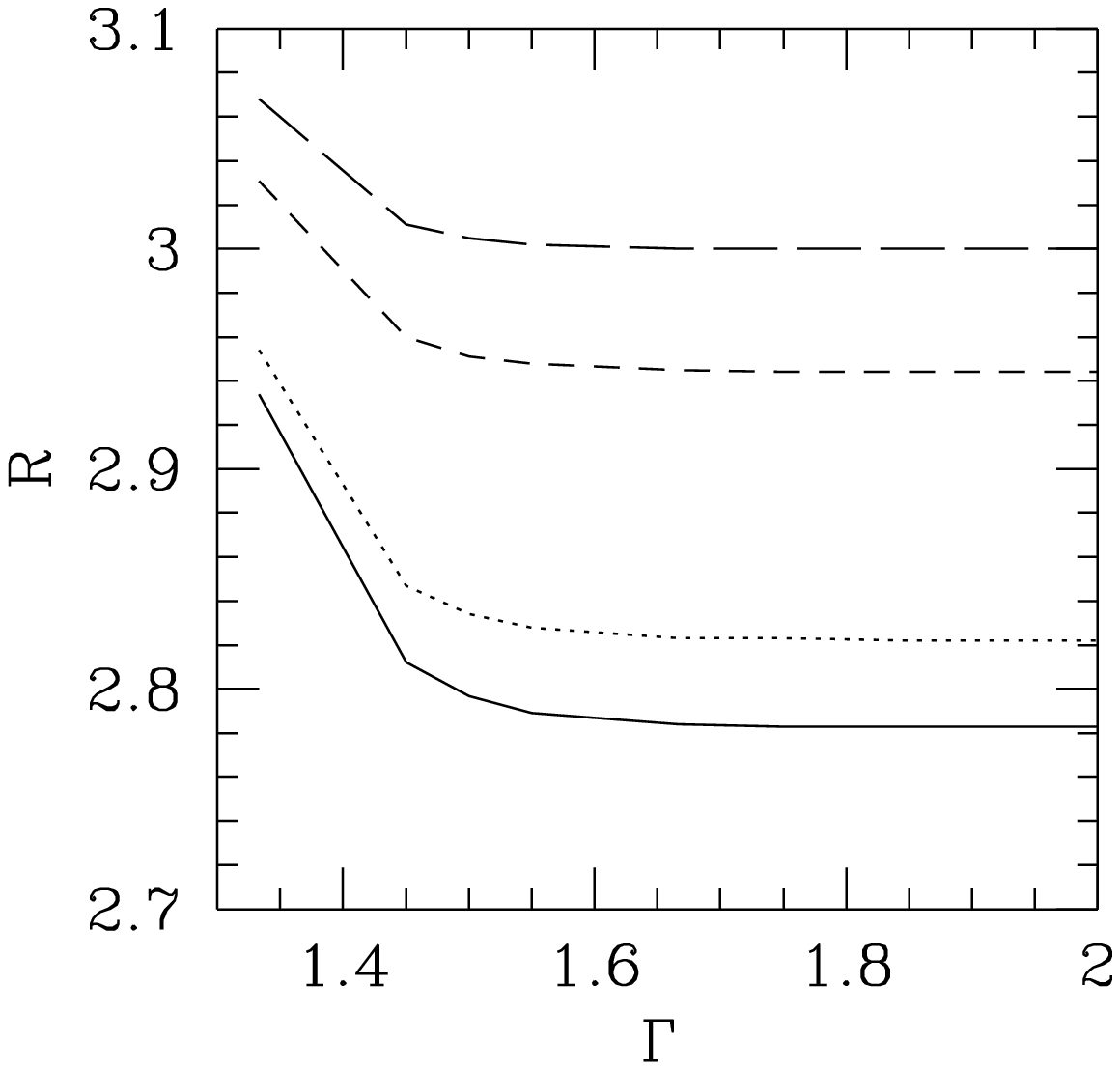,height=10truecm,width=10truecm,angle=0}}}
\noindent {\small {\bf Fig. 8} : Variation of the location of centrifugal reversal
as a function of $\Gamma$ for Kerr parameter (a) $a=1$ (solid curve), (b) $a=0.9$ (dotted curve),
(c) $a=0.5$ (dashed curve) and (d) $a=0$ (long-dashed curve).
}
\end{figure}
 
If we consider the Coriolis force as given by (\ref{coi}), it is seen immediately that
while for static space time it is zero, as it should be, in Kerr space time only the
radial component is non zero in Boyer-Lindquist coordinate system and is given by
 
\begin{eqnarray}
\label{cor}
(Co)_r=-\frac{2 a m\gamma^2 V^\phi}{\alpha r} \left(\frac{3r^2+a^2}{r^3+a^2 r+2 m a^2}\right)
\end{eqnarray}
 
\noindent which matches exactly with the expression for the test particle case. This in fact
is to be expected as Coriolis force does in no way depend upon the radial velocity, in the present context.
 
In conclusion we can say that for a fluid in curved space time, the centrifugal force
reversal occurs only through the dominance of azimuthal velocity over the radial
velocity. For Kerr black hole, due to the effect of inertial frame dragging, which always
goes to increase the effective angular velocity, the reversal can occur even when the
pressure of the fluid is zero. On the other hand when pressure is non zero, the
gradient of pressure seems to bring down the effect of radial velocity and allow
the fluid to have larger azimuthal velocity and thus lead to centrifugal reversal
both in radial and azimuthal directions. The above discussion, however did not
take into account any viscosity, which if present may affect the azimuthal velocity
through the transfer of angular momentum and thus increase the effect of radial
velocity. It would be necessary to extend these studies by including the
viscous forces and analyse the effect on centrifugal reversal.

\end{document}